\documentclass[twocolumn,showpacs,showkeys,preprintnumbers,amsmath,amssymb]{revtex4}
\usepackage{graphicx}% Include figure files
\begin{document}

%\preprint{astro-ph/090200}

\title{Scaling Laws in High-Energy Inverse Compton Scattering}
\author{Satoshi Nozawa}
 \email{snozawa@josai.ac.jp}
\affiliation{
Josai Junior College, 1-1 Keyakidai, Sakado-shi, Saitama, 350-0295,
Japan}

\author{Yasuharu Kohyama and Naoki Itoh}
\affiliation{
Department of Physics, Sophia University, 7-1 Kioi-cho, Chiyoda-ku,
Tokyo, 102-8554, Japan}

\date{\today}% It is always \today, today,

\begin{abstract}
  Based upon the rate equations for the photon distribution function obtained in the previous paper, we study the inverse Compton scattering process for high-energy nonthermal electrons.  Assuming the power-law electron distribution, we find a scaling law in the probability distribution function $P_{1}(s)$, where the peak height and peak position depend only on the power index parameter.  We solved the rate equation analytically.  It is found that the spectral intensity function also has the scaling law, where the peak height and peak position depend only on the power index parameter.  The present study will be particularly important to the analysis of the X-ray and gamma-ray emission models from various astrophysical objects such as radio galaxies and supernova remnants.
\end{abstract}



\maketitle

\section{Introduction}

  The inverse Compton scattering is one of the most fundamental reactions which have variety of applications to astrophysics and cosmology.  They are, for example, the Sunyaev-Zeldovich (SZ) effects\cite{suny72} for clusters of galaxies (CG), cosmic-ray emission from radio galaxies\cite{blun06} and clusters of galaxies\cite{sara99}, and radio to gamma-ray emission from supernova remnants\cite{bari99, laze04}.  Therefore, theoretical studies on the inverse Compton scattering have been done quite extensively for the last forty years, starting from the works by Jones\cite{jone68}, and Blumenthal and Gould\cite{blum70} to the recent works, for example, by Fargion\cite{farg97}, Colafrancesco\cite{cola08, cola09}, and Petruk\cite{petr09}.

  In particular, remarkable progress has been made in theoretical studies for the SZ effects for CG.  Wright\cite{wrig79} and Rephaeli\cite{reph95} calculated the photon frequency redistribution function in the electron rest frame, which is called as the radiative transfer method.  On the other hand, Challinor and Lasenby\cite{chal98} and Itoh, Kohyama, and Nozawa\cite{itoh98} solved the relativistically covariant Boltzmann collisional equation for the photon distribution function, which is called the covariant formalism.  Although the two are very different approaches, the obtained results for the SZ effect agreed extremely well.  This has been a longstanding puzzle in the field of the relativistic study of the SZ effect for the last ten years.  Very recently, however, Nozawa and Kohyama\cite{noza09a} (denoted NK hereafter) showed that the two formalisms were indeed mathematically equivalent in the approximation of the Thomson limit.  This explained the reason why the two different approaches produced same results for the SZ effect even in the relativistic energies for electrons.

  In the present paper, we extend the formalism obtained by NK to the case of high-energy electrons.  This extension will be particularly interesting for the analysis of X-ray and gamma-ray emissions, for example, from radio galaxies\cite{blun06} and supernova remnants\cite{bari99, laze04}, where the inverse Compton scattering of the CMB photons off nonthermal high-energy electrons plays an essential role.  In the present approach, we push analytic techniques as much as possible in order to obtain analytic solutions.  In contrast to the direct numerical calculation, the present approach will have an advantage that one may reveal essential physics properties behind the numerical results.  In the present paper, under a specific condition for the electron distribution which is typically realized, we will show that a universal scaling law is valid for the spectral intensity function.

  The present paper is organized as follows:  Starting from the rate equations derived in the NK paper, we derive in Sec.~II the analytic expressions for the redistribution function $P(s,\gamma)$ and probability distribution function $P_{1}(s)$.  Assuming the power-law electron distribution, we show that $P_{1}(s)$ has a scaling law, where the peak height and peak position depend only on the power index parameter.  We calculated the rate equation and obtained the analytic expression for the spectral intensity function $dI(X)/d\tau$.  We show that $dI(X)/d\tau$ also has the scaling law, where the peak height and peak position depend only on the power index parameter.  In Sec.~III we apply the scaling law to the observation of the spectral intensity in the X-ray and gamma-ray energy regions.  Finally, concluding remarks are given in Sec.~IV.

\section{High-energy Inverse Compton scattering}

\subsection{Rate Equations in Thomson Approximation}

  In the NK paper, it was shown that the covariant formalism\cite{itoh98} and radiative transfer method\cite{wrig79} were mathematically equivalent in the following (Thomson) approximation:
\begin{eqnarray}
&&\hspace{-10mm}
\gamma \frac{\omega}{m} \ll 1  \, ,
\label{eq2a-1}   \\
&&\hspace{-10mm}
\gamma = \frac{1}{\sqrt{1-\beta^{2}}}  \, ,
\label{eq2a-2}
\end{eqnarray}
where $\omega$ is the photon energy, $\gamma$ is the Lorentz factor, and $\beta$ and $m$ are the velocity and rest mass of the electron, respectively.  Throughout this paper, we use the natural unit $\hbar = c = 1$, unless otherwise stated explicitly.  For the cosmic microwave background (CMB) photons ($k_{B}T_{CMB}=2.348 \times 10^{-4}$eV), $\omega < 5 \times 10^{-3}$eV is well satisfied.  Then $\omega/m < 1 \times 10^{-8}$, which implies $\gamma \ll 10^{8}$.  Therefore as far as the CMB photons are concerned, Eq.~(\ref{eq2a-1}) is fully valid from non-relativistic electrons to extreme-relativistic electrons of the order of TeV region.

  The rate equations for the photon distribution function $n(x)$ and spectral intensity function $I(x)$ were derived in the NK paper under the assumption of Eq.~(\ref{eq2a-1}).  Here, $x=\omega/k_{B}T_{CMB}$ is the photon energy in units of the thermal energy of the CMB.   We recall the results here to make the present paper more self-contained.  They are given as follows\cite{noza09a, noza09b}:
\begin{eqnarray}
&&\hspace{-10mm}
\frac{\partial n(x)}{\partial \tau}
 = \int_{-\infty}^{\infty}ds P_1(s)
\left[n(e^sx)- n(x)\right] \, ,
\label{eq2a-3}   \\
&&\hspace{-10mm}
\frac{\partial I(x)}{\partial \tau}
 = \int_{-\infty}^{\infty}ds
{P}_1(s) \left[I(e^{-s}x)- I(x)\right]  \, ,
\label{eq2a-4}  \\
&&\hspace{+12mm}
\tau  =  n_e\sigma_T t \, , 
\label{eq2a-5}
\end{eqnarray}
where $I(x)=I_{0}x^{3}n(x)$, $I_{0}=(k_{B}T_{CMB})^{3}/2\pi^{2}$, $n_{e}$ is the electron number density, $\sigma_{T}$ is the Thomson scattering cross section.  In Eqs.~(\ref{eq2a-3}) and (\ref{eq2a-4}), $P_{1}(s)$ is the probability distribution function for the photon of a frequency shift $s$, which is defined by $e^{s}=x^{\prime}/x$,
\begin{eqnarray}
&&\hspace{-10mm}
P_1(s) = \int_{\beta_{min}}^{1}d\beta\beta^2\gamma^5 p_e(E)P(s,\beta)  \, ,
\label{eq2a-6}  \\
&&\hspace{-10mm}
P(s,\beta) = \frac{e^{s}}{2\beta\gamma^4}
\int_{\mu_1(s)}^{\mu_2(s)}d\mu_0 \frac{1}{(1-\beta\mu_0)^2} f\left(\mu_0, \mu_0^{\prime} \right)   \, ,
\label{eq2a-7}  \\
&&\hspace{-10mm}
f(\mu_0,\mu_0^{\prime}) = \frac{3}{8}\left[
  1 + \mu_0^2\mu_0^{\prime 2}+\frac{1}{2}(1-\mu_0^2)(1-\mu_0^{\prime 2})
\right]  \, ,
\label{eq2a-8}
\end{eqnarray}
where $p_{e}(E)$ is the electron distribution function of a momentum $p$ which is normalized by $\int_{0}^{\infty} dp p^{2} p_{e}(E)/m^{3}=1$.  Variables appearing in Eqs.~(\ref{eq2a-6}) -- (\ref{eq2a-8}) are summarized as follows:
\begin{eqnarray}
&&\hspace{-10mm}
\beta_{min} = (1-e^{-|s|})/(1+e^{-|s|})  \, ,
\label{eq2a-9} \\
&&\hspace{-10mm}
\mu_{0}^{\prime} = [1-e^s(1-\beta\mu_0)]/\beta  \, ,
\label{eq2a-10}  \\
&&\hspace{-10mm}
\mu_1(s) = \left\{
\begin{array}{ll}
-1 &\quad  {\rm for} \, \, \, s \leq 0 \\
{[1-e^{-s}(1+\beta)]/\beta} &\quad {\rm for} \, \, \, s > 0
\end{array}
\right.  \, ,
\label{eq2a-11} \\
&&\hspace{-10mm}
\mu_2(s) = \left\{
\begin{array}{ll}
{[1-e^{-s}(1-\beta)]/\beta} &\quad {\rm for} \, \, \, s < 0 \\
1 &\quad  {\rm for} \, \, \, s \geq 0 
\end{array}
\right. \, .
\label{eq2a-12}
\end{eqnarray}

  The total probabilities for $P(s,\beta)$ and $P_{1}(s)$ are given by
\begin{eqnarray}
&&\hspace{-20mm}
\int_{-\lambda_{\beta}}^{+\lambda_{\beta}}ds P(s, \beta) = 1
\label{eq2a-13}  \, , \\
&&\hspace{-15mm}
\int_{-\infty}^{\infty}ds P_1(s) = 1
\label{eq2a-14}  \, ,
\end{eqnarray}
where
\begin{eqnarray}
&&\hspace{-10mm}
\lambda_{\beta} = \ln \left(\frac{1+\beta}{1-\beta} \right)
\label{eq2a-15}  \, .
\end{eqnarray}
It should be noted that the following useful relations
\begin{eqnarray}
&&\hspace{-10mm}
P(s, \beta)e^{-3s} = P(-s,\beta)
\label{eq2a-16}  \, , \\
&&\hspace{-7mm}
P_{1}(s)e^{-3s} = P_{1}(-s)
\label{eq2a-17}
\end{eqnarray}
are valid.

\subsection{$P(s, \beta)$ for Extreme-Relativistic Electrons}

 In this section, we derive the analytic expression of the frequency redistribution function $P(s, \beta)$ for extreme-relativistic electrons.  In Eq.~(\ref{eq2a-7}), the integral of $\mu_{0}$ can be done analytically.  One obtains as follows: for $s<0$,
\begin{eqnarray}
&&\hspace{-10mm}
P(s,\beta) = \frac{3}{32\beta^2\gamma^4}
\left[
-C_1(\beta)-C_2(\beta)e^s+C_3(\beta)e^{2s}
\right.
\nonumber   \\
&&\hspace{+5mm}
\left.
+C_4(\beta)(\lambda_{\beta}+s)(e^s+e^{2s})
+C_1(\beta)e^{3s}
\right]  \, ,
\label{eq2b-1}
\end{eqnarray}
and for $s\ge0$,
\begin{eqnarray}
&&\hspace{-10mm}
P(s,\beta) = \frac{3}{32\beta^2\gamma^4}
\left[
C_1(\beta)+C_3(\beta)e^s-C_2(\beta)e^{2s}
\right.
\nonumber   \\
&&\hspace{+5mm}
\left.
+C_4(\beta)(\lambda_{\beta}-s)(e^s+e^{2s})
-C_1(\beta)e^{3s}
\right]  \, ,
\label{eq2b-2}
\end{eqnarray}
where the coefficients are
\begin{eqnarray}
&&\hspace{-10mm}
C_1(\beta) = \frac{1}{\beta^4\gamma^2}
\label{eq2b-3} \, , \\
&&\hspace{-10mm}
C_2(\beta) = \frac{1}{\beta^{4}(1+\beta)} \left(4\beta^4-\beta^3-13\beta^2-3\beta+9\right)
\label{eq2b-4} \, , \\
&&\hspace{-10mm}
C_3(\beta) =  \frac{1}{\beta^{4}(1-\beta)} \left(4\beta^4+\beta^3-13\beta^2+3\beta+9\right)
\label{eq2b-5} \, , \\
&&\hspace{-10mm}
C_4(\beta) = \frac{2}{\beta^4}\left(\beta^2-3\right)
\label{eq2b-6} \, .
\end{eqnarray}
Note that Eqs.~(\ref{eq2b-1}) and (\ref{eq2b-2}) satisfy the relation of Eq.~(\ref{eq2a-16}).  It should be also noted that Eqs.~(\ref{eq2b-1}) and (\ref{eq2b-2}) agree with Eqs.~(23a) and (23b) of Fargion et al.\cite{farg97}, respectively.

  Now let us consider the case for electrons of extreme-relativistic energies $E$ $(=\gamma mc^{2}) \gg mc^{2}$.  Thus, $\gamma \gg 1$ and $\beta \approx 1$ are assumed.  In this approximation, Eqs.~(\ref{eq2b-1}) and (\ref{eq2b-2}) are written as follows:  for $s<0$,
\begin{eqnarray}
&&\hspace{-10mm}
P(s,\gamma) = \frac{3}{32\gamma^4}
\left[-\frac{1}{\gamma^2}+2e^s+8\gamma^2e^{2s}  \right.
\nonumber  \\
&&\hspace{+10mm}
\left. -4\left(\lambda_{\gamma}+s\right)\left(e^s+e^{2s}\right)
+\frac{1}{\gamma^2}e^{3s}\right]  \, ,
\label{eq2b-7}
\end{eqnarray}
and for $s\ge0$,
\begin{eqnarray}
&&\hspace{-10mm}
P(s,\gamma) = \frac{3}{32\gamma^4}\left[\frac{1}{\gamma^2}+8\gamma^2e^s
+2e^{2s}  \right.
\nonumber  \\
&&\hspace{+10mm}
\left. -4\left(\lambda_{\gamma}-s\right)\left(e^s+e^{2s}\right)
-\frac{1}{\gamma^2}e^{3s}\right]  \, ,
\label{eq2b-8}  \\
&&\hspace{-3mm}
\lambda_{\gamma} = 2 {\rm ln}(2 \gamma)
\label{eq2b-9} \, ,
\end{eqnarray}
where the expression $P(s,\gamma)$ was used instead of $P(s,\beta)$.  Equations (\ref{eq2b-7}) and (\ref{eq2b-8}) can be integrated analytically.  One obtains as follows: for $s<0$,
\begin{eqnarray}
&&\hspace{-10mm}
\int_{-\lambda_{\gamma}}^0ds P(s,\gamma) = O\left(\frac{1}{\gamma^2}\right)
\label{eq2b-10} \, ,
\end{eqnarray}
and for $s\ge0$,
\begin{eqnarray}
&&\hspace{-5mm}
\int_{0}^{\lambda_{\gamma}}ds P(s,\gamma) = 1+ O\left(\frac{1}{\gamma^2}\right)
\label{eq2b-11} \, .
\end{eqnarray}
In Eq.~(\ref{eq2b-11}), the terms contributing to $O\left(1/\gamma^2\right)$ are the higher-order terms.  Therefore one can eliminate the corresponding terms from Eq.~(\ref{eq2b-8}), which gives the new expression for $s\ge0$.  Then the new expression for $s<0$ is obtained with the relation of Eq.~(\ref{eq2a-16}).  Therefore, the total probability
\begin{eqnarray}
&&\hspace{-10mm}
\int_{-\lambda_{\gamma}}^{+\lambda_{\gamma}}ds P(s, \gamma) = 1 + O\left(\frac{1}{\gamma^{2}}\right)
\label{eq2b-12}
\end{eqnarray}
is satisfied for $P(s,\gamma)$.  The explicit forms are as follows: for $s<0$,
\begin{eqnarray}
&&\hspace{-15mm}
P(s,\gamma) = \frac{3}{32\gamma^4}
\Bigl[-\frac{1}{\gamma^2}+2e^s+8\gamma^2e^{2s}
\nonumber  \\
&&\hspace{+15mm}
-4\left(\lambda_{\gamma}+s\right)e^s \Bigr]
\label{eq2b-13} \, ,
\end{eqnarray}
and for $s\ge0$,
\begin{eqnarray}
&&\hspace{-10mm}
P(s,\gamma) = \frac{3}{32\gamma^4}\Bigl[ 8\gamma^2e^s
+2e^{2s}
\nonumber  \\
&&\hspace{+15mm}
-4\left(\lambda_{\gamma}-s\right)e^{2s}
-\frac{1}{\gamma^2}e^{3s} \Bigr]
\label{eq2b-14} \, .
\end{eqnarray}

  Let us now compare the present results with the literature.  It is straightforward to show that Eqs.~(\ref{eq2b-13}) and (\ref{eq2b-14}) are equivalent to Eqs.~(38) and (40) of Jones\cite{jone68}, respectively.  We show the equivalence between the present formalism and Jones's formalism in Appendix A.  It should be also mentioned that Eqs.~(24a) and (24b) of Fargion et al.\cite{farg97} differ from our Eqs.~(\ref{eq2b-13}) and (\ref{eq2b-14}).  The difference comes from $O(1/\gamma^{2})$ terms as mentioned in their paper.
  
  Before closing this subsection, it should be also noted the following:  In the present formalism, the cases $s\ge0$ and $s<0$ correspond to the Compton scattering and inverse Compton scattering, respectively.  This is because of the definition $x = e^{-s} x^{\prime}$, where $x^{\prime}$ and $x$ are the energies (in units of $k_{B}T_{CMB}$) of initial and final photons, respectively.  Equations (\ref{eq2b-10}) and (\ref{eq2b-11}) suggest that probability distribution for the CMB photon scattering by high-energy electrons is dominated by the Compton scattering process instead of the inverse Compton scattering process.

\subsection{Scaling Law of $P_{1}(s)$ for Nonthermal Electrons}

  In order to proceed calculation for practical applications, let us specify the electron distribution function.  High-energy electrons in the supernova remnants and active galactic nuclei, for example, are most likely nonthermal.  It is standard to describe the nonthermal distribution in terms of the power-law distribution function of three parameters:
\begin{eqnarray}
&&\hspace{-10mm}
p_{e}(\gamma) = \left\{
\begin{array}{ll}
 N_{\gamma} \, \gamma^{-\sigma} \, , &  \, \, \, \gamma_{min} \leq \gamma \leq \gamma_{max} \\
   0  \, , &  \, \, \, {\rm elsewhere}
\end{array}
\right.  \, ,
\label{eq2c-1}
\end{eqnarray}
where $\gamma$ is the Lorentz factor and $N_{\gamma}$ is the normalization constant.  In Eq.~(\ref{eq2c-1}), $\sigma$ is the power index parameter, $\gamma_{min}$ and $\gamma_{max}$ are parameters of minimum and maximum values for $\gamma$, respectively.  Then, Eq.~(\ref{eq2a-6}) can be reexpressed as follows: for $s<0$,
\begin{eqnarray}
&&\hspace{-10mm}
P_1(s) = \int_{\max(\gamma_{min},e^{-s/2}/2)}^{\gamma_{max}}d\gamma p_e(\gamma) P(s,\gamma)  \, ,
\label{eq2c-2}
\end{eqnarray}
where $P(s,\gamma)$ is given by Eq.~(\ref{eq2b-13}), and for $s\ge0$,
\begin{eqnarray}
&&\hspace{-10mm}
P_1(s) = \int_{\max(\gamma_{min},e^{s/2}/2)}^{\gamma_{max}}d\gamma p_e(\gamma)P(s,\gamma)  \, ,
\label{eq2c-3}
\end{eqnarray}
where $P(s,\gamma)$ is given by Eq.~(\ref{eq2b-14}).  In deriving Eqs.~(\ref{eq2c-2}) and (\ref{eq2c-3}), $\beta \approx 1$ was assumed, and the phase space factor $\gamma^{2}$ was absorbed, for simplicity, by the power index $\sigma$ in $p_e(\gamma)$.  

  In the case of the power-law distribution of Eq.~(\ref{eq2c-1}), equations (\ref{eq2c-2}) and (\ref{eq2c-3}) can be integrated analytically.  The explicit forms are given as follows: for $-2\ln 2\gamma_{max}<s <-2\ln 2\gamma_{min} $,
\begin{eqnarray}
&&\hspace{-5mm}
P_1(s) = \frac{3}{32}N_{\gamma} \left\{
-\frac{1}{\sigma+5}
\left(2^{\sigma+5}e^{(\sigma+5)s/2}-\frac{1}{\gamma_{max}^{\sigma+5}}\right)
\right.
\nonumber \\
&&\hspace{4mm}
+\frac{2}{\sigma+3}\left[
\frac{\sigma-1}{\sigma+3}2^{\sigma+3}
e^{(\sigma+3)s/2}
\right.
\nonumber \\
&&\hspace{15mm}
\left.
-\frac{1}{\gamma_{max}^{\sigma+3}}
\left(\frac{\sigma-1}{\sigma+3}-2s-4\ln 2\gamma_{max}\right)
\right]e^s
\nonumber \\
&&\hspace{4mm}
\left.
+\frac{8}{\sigma+1}
\left(
2^{\sigma+1}e^{(\sigma+1)s/2}
-\frac{1}{\gamma_{max}^{\sigma+1}}\right)
e^{2s}
\right\}
\label{eq2c-4}  \, ,
\end{eqnarray}
for $-2\ln 2\gamma_{min}<s<0$,
\begin{eqnarray}
&&\hspace{-5mm}
P_1(s) = \frac{3}{32}N_{\gamma} \left\{
-\frac{1}{\sigma+5}
\left(\frac{1}{\gamma_{min}^{\sigma+5}}-\frac{1}{\gamma_{max}^{\sigma+5}}\right)
\right.
\nonumber \\
&&\hspace{4mm}
+\frac{2}{\sigma+3}\left[
\frac{1}{\gamma_{min}^{\sigma+3}}
\left(\frac{\sigma-1}{\sigma+3}-2s-4\ln 2\gamma_{min}\right)
\right.
\nonumber \\
&&\hspace{15mm}
\left.
-\frac{1}{\gamma_{max}^{\sigma+3}}
\left(\frac{\sigma-1}{\sigma+3}-2s-4\ln 2\gamma_{max}\right)
\right]e^s
\nonumber \\
&&\hspace{4mm}
\left.
+\frac{8}{\sigma+1}
\left(\frac{1}{\gamma_{min}^{\sigma+1}}-\frac{1}{\gamma_{max}^{\sigma+1}}\right)
e^{2s}
\right\}
\label{eq2c-5}  \, ,
\end{eqnarray}
for $0 < s < 2\ln 2\gamma_{min}$,
\begin{eqnarray}
&&\hspace{-5mm}
P_1(s) = \frac{3}{32}N_{\gamma} \left\{
-\frac{1}{\sigma+5}
\left(\frac{1}{\gamma_{min}^{\sigma+5}}-\frac{1}{\gamma_{max}^{\sigma+5}}\right)
e^{3s}
\right.
\nonumber \\
&&\hspace{4mm}
+\frac{2}{\sigma+3}\left[
\frac{1}{\gamma_{min}^{\sigma+3}}
\left(\frac{\sigma-1}{\sigma+3}+2s-4\ln 2\gamma_{min}\right)
\right.
\nonumber \\
&&\hspace{15mm}
\left.
-\frac{1}{\gamma_{max}^{\sigma+3}}
\left(\frac{\sigma-1}{\sigma+3}+2s-4\ln 2\gamma_{max}\right)
\right]e^{2s}
\nonumber \\
&&\hspace{4mm}
\left.
+\frac{8}{\sigma+1}
\left(\frac{1}{\gamma_{min}^{\sigma+1}}-\frac{1}{\gamma_{max}^{\sigma+1}}\right)
e^{s}
\right\}
\label{eq2c-6}  \, ,
\end{eqnarray}
and for $2\ln 2\gamma_{min}<s <2\ln 2\gamma_{max} $,
\begin{eqnarray}
&&\hspace{-5mm}
P_1(s) = \frac{3}{32}N_{\gamma} \left\{
-\frac{1}{\sigma+5}
\left(2^{\sigma+5}e^{-(\sigma+5)s/2}-\frac{1}{\gamma_{max}^{\sigma+5}}\right)
e^{3s}
\right.
\nonumber \\
&&\hspace{4mm}
+\frac{2}{\sigma+3}\left[
\frac{\sigma-1}{\sigma+3}
2^{\sigma+3}e^{-(\sigma+3)s/2}
\right.
\nonumber \\
&&\hspace{15mm}
\left.
-\frac{1}{\gamma_{max}^{\sigma+3}}
\left(\frac{\sigma-1}{\sigma+3}+2s-4\ln 2\gamma_{max}\right)
\right]e^{2s}
\nonumber \\
&&\hspace{4mm}
\left.
+\frac{8}{\sigma+1}
\left(
2^{\sigma+1}e^{-(\sigma+1)s/2}
-\frac{1}{\gamma_{max}^{\sigma+1}}\right)
e^{s}
\right\}
\label{eq2c-7}  \, .
\end{eqnarray}
It should be noted that the normalization constant is given by
\begin{eqnarray}
&&\hspace{-10mm}
N_{\gamma} = (\sigma-1)\gamma_{min}^{\sigma-1}
\label{eq2c-8}
\end{eqnarray}
for the case $\gamma_{\max} \to \infty$.

  Let us now introduce new functions $P_{C}(s,R)$ and $P^{\prime}_{IC}(s,R)$ in order to express Eqs.~(\ref{eq2c-4})--(\ref{eq2c-7}) in unified forms, where $R=\gamma_{min}/\gamma_{max}$.  Here, $C$ and $IC$ denote the Compton scattering and Inverse Compton scattering, respectively.  First, we define $P_{C}(s,R)$ as follows: for $-2\ln 2\gamma_{min} < s < 0$,
\begin{eqnarray}
&&\hspace{-5mm}
P_{C}(s,R) = 3\frac{\sigma-1}{1-R^{\sigma-1}}\left\{
-\frac{2}{\sigma+5}
\left(1-R^{\sigma+5}\right)
e^{3s}
\right.
\nonumber \\
&&\hspace{15mm}
+\frac{1}{\sigma+3}\left[
\frac{\sigma-1}{\sigma+3}+2s
\right.
\nonumber \\
&&\hspace{14mm}
\left.
-R^{\sigma+3}
\left(\frac{\sigma-1}{\sigma+3}+2s+4\ln R\right)
\right]e^{2s}
\nonumber \\
&&\hspace{14mm}
\left.
+\frac{1}{\sigma+1}
\left(1-R^{\sigma+1}\right)
e^{s}
\right\}
\label{eq2c-9} \, ,
\end{eqnarray}
and for $0<s <2\ln (\gamma_{max}/\gamma_{min}) $,
\begin{eqnarray}
&&\hspace{-5mm}
P_{C}(s,R) = 3\frac{\sigma-1}{1-R^{\sigma-1}}\left\{
-\frac{2}{\sigma+5}
\left(e^{-(\sigma+5)s/2}-R^{\sigma+5}\right)
e^{3s}
\right.
\nonumber \\
&&\hspace{11mm}
+\frac{1}{\sigma+3}\left[
\frac{\sigma-1}{\sigma+3}
e^{-(\sigma+3)s/2}
\right.
\nonumber \\
&&\hspace{11mm}
\left.
-R^{\sigma+3}
\left(\frac{\sigma-1}{\sigma+3}+2s+4\ln R\right)
\right]e^{2s}
\nonumber \\
&&\hspace{11mm}
\left.
+\frac{1}{\sigma+1}
\left(
e^{-(\sigma+1)s/2}
-R^{\sigma+1}\right)
e^{s}
\right\}
\label{eq2c-10} \, .
\end{eqnarray}
Similarly, $P^{\prime}_{IC}(s,R)$ is for $-2\ln (\gamma_{max}/\gamma_{min})< s < 0 $,
\begin{eqnarray}
&&\hspace{-5mm}
P^{\prime}_{IC}(s,R) = 3\frac{\sigma-1}{1-R^{\sigma-1}}\left\{
-\frac{2}{\sigma+5}
\left(e^{(\sigma+5)s/2}-R^{\sigma+5}\right)
\right.
\nonumber \\
&&\hspace{13mm}
+\frac{1}{\sigma+3}\left[
\frac{\sigma-1}{\sigma+3}
e^{(\sigma+3)s/2}
\right.
\nonumber \\
&&\hspace{13mm}
\left.
-R^{\sigma+3}
\left(\frac{\sigma-1}{\sigma+3}-2s+4\ln R\right)
\right]e^{s}
\nonumber \\
&&\hspace{13mm}
\left.
+\frac{1}{\sigma+1}
\left(
e^{(\sigma+1)s/2}
-R^{\sigma+1}\right)
e^{2s}
\right\}
\label{eq2c-11} \, ,
\end{eqnarray}
and for $0<s <2\ln 2\gamma_{min} $,
\begin{eqnarray}
&&\hspace{-10mm}
P^{\prime}_{IC}(s,R) = 3\frac{\sigma-1}{1-R^{\sigma-1}}\left\{
-\frac{2}{\sigma+5}
\left(1-R^{\sigma+5}\right)
\right.
\nonumber \\
&&\hspace{10mm}
+\frac{1}{\sigma+3}\left[
\frac{\sigma-1}{\sigma+3}-2s
\right.
\nonumber \\
&&\hspace{10mm}
\left.
-R^{\sigma+3}
\left(\frac{\sigma-1}{\sigma+3}-2s+4\ln R\right)
\right]e^{s}
\nonumber \\
&&\hspace{10mm}
\left.
+\frac{1}{\sigma+1}
\left(1-R^{\sigma+1}\right)
e^{2s}
\right\}
\label{eq2c-12} \, .
\end{eqnarray}
It is straightforward to show that
\begin{eqnarray}
&&\hspace{-10mm}
P_{C}(s,R) e^{-3s} = P^{\prime}_{IC}(-s,R)
\label{eq2c-13}
\end{eqnarray}
is satisfied by Eqs.~(\ref{eq2c-9})--(\ref{eq2c-12}).

  Comparing Eqs.~(\ref{eq2c-4})--(\ref{eq2c-7}) with Eqs.~(\ref{eq2c-9})--(\ref{eq2c-12}), the probability distribution function $P_{1}(s)$ is described as follows:
\begin{eqnarray}
&&\hspace{-13mm}
P_{1}(s) = 
\left\{
\begin{array}{ll}
P_{IC}(s+2\ln 2\gamma_{min},R) &   {\rm for \,} s < 0 \\
  \\
P_{C}(s-2\ln 2\gamma_{min},R)  & {\rm for \,} s \ge 0
\end{array}
\right.
\label{eq2c-14} \, ,
\end{eqnarray}
where
\begin{eqnarray}
&&\hspace{-10mm}
P_{IC}(s,R) \equiv \frac{1}{64\gamma_{min}^6} P^{\prime}_{IC}(s,R)
\label{eq2c-15}  \, .
\end{eqnarray}

  Let us now consider the case $R \equiv \gamma_{min}/\gamma_{max} \ll 1$.  We fix $\gamma_{max}$ = 10$^{8}$ throughout the paper.  In Fig.~1(a), we plot $P_{1}(s)$ as a function of $s$ for a typical value $\sigma=2.5$.  The solid curve, dash-dotted curve, dashed curve, and dotted curve correspond to $\gamma_{min}$ = 10, 10$^{2}$, 10$^{3}$, and 10$^{4}$, respectively.  It can be seen that the height of $P_{1}(s)$ is independent of $\gamma_{min}$.  In Fig.~1(b), we plot the same curves as a function of new variable $s_{C}$ which is defined by
\begin{eqnarray}
&&\hspace{-10mm}
s_{C}= s - 2 \ln 2 \gamma_{min}
\label{eq2c-16}  \, .
\end{eqnarray}
In Fig.~1(b) the four curves are totally indistinguishable, which exhibits a scaling law for $P_{1}(s)$.  The reason for this scaling law is as below.  For large $\gamma_{min} \gg 1$, as shown by Figs.~1(a), 1(b), and Eqs.~(\ref{eq2c-14}) and (\ref{eq2c-15}), the probability distribution function $P_{1}(s)$ is dominated by $P_{C}(s_{C},0)$, i.e. by the Compton scattering process.

%______________________________________________ Fig.\ 1
\begin{figure}
\begin{center}
\includegraphics[angle=0,width=0.48\textwidth]{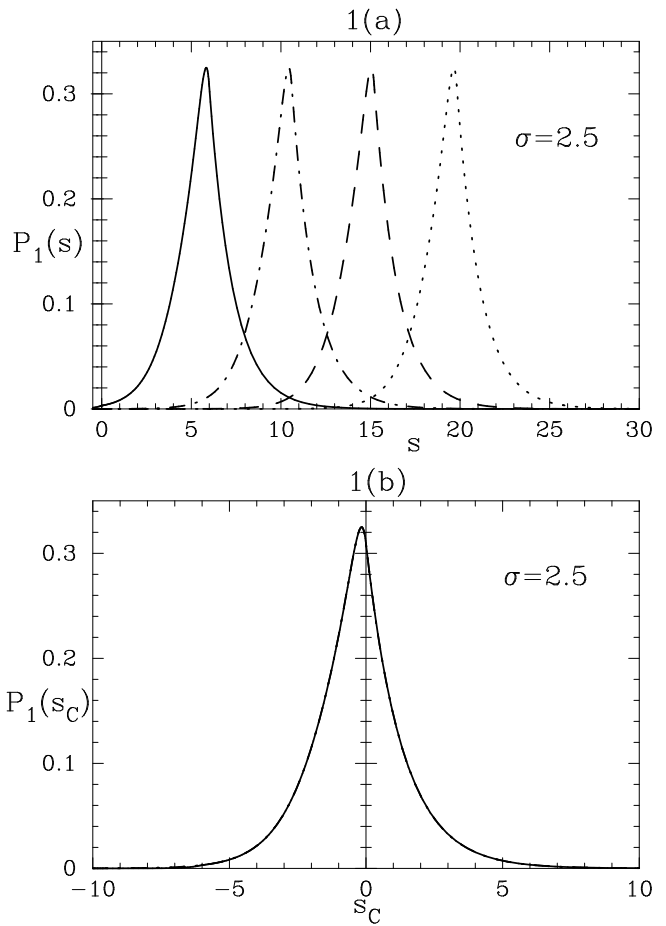}
\end{center}
\caption{Plotting of $P_{1}(s)$ and $P_{1}(s_{C})$ for $\sigma=2.5$.  Figures 1(a) and 1(b) are $P_{1}(s)$ and $P_{1}(s_{C})$, respectively.  The solid curve, dash-dotted curve, dashed curve, and dotted curve correspond to $\gamma_{min}$ = 10, 10$^{2}$, 10$^{3}$, and 10$^{4}$, respectively.}
\end{figure}

  Before closing this subsection, we study the $\sigma$-dependences on the peak position $s_{peak}$ and peak height $P_{1}(s_{peak})$.  As shown in Figs.\ 1(a) and 1(b), the $\gamma_{min}$-dependence of $P_{1}(s)$ is described by Eq.\ (\ref{eq2c-16}), namely, $s$ = $s_{C} + 2 \ln 2 \gamma_{min}$.  Therefore, we define the peak position by
\begin{eqnarray}
&&\hspace{-10mm}
s_{peak} =  s(\sigma) + 2 \ln 2\gamma_{min}
\label{eq2c-18} \, ,
\end{eqnarray}
where $s(\sigma)$ depends only on $\sigma$.  The peak position is calculated by solving the equation
\begin{eqnarray}
&&\hspace{-10mm}
\left.
\frac{ \partial  P_{1}(s)}{ \partial s} \right|_{s_{peak}} = 0
\label{eq2c-19}  \, .
\end{eqnarray}
  The analytic expressions for $s(\sigma)$ in the first-order and third-order approximations are given as follows:
\begin{eqnarray}
&&\hspace{-10mm}
s_{1st}(\sigma) = -\frac{(\sigma-1)(\sigma^{2}+ 4\sigma + 11)}{5 \sigma^{3} + 23\sigma^{2} + 51 \sigma +17}
\label{eq2c-20}  \, , \\
&&\hspace{-10mm}
s_{3rd}(\sigma) = -\frac{1}{2(4 \sigma^{2} + 21\sigma + 29)} \Biggl[ 3\sigma^{2}+ 14\sigma + 19   \\
\nonumber  \\
&&\hspace{8mm}
+ \left( \frac{ \sqrt{(\sigma+1)^{2} A^{3} + B^{2}} + B }{ \sigma+1} \right)^{1/3} \nonumber  \\
&&\hspace{8mm}
\left. - \left( \frac{ \sqrt{(\sigma+1)^{2} A^{3} + B^{2}} - B }{ \sigma+1} \right)^{1/3} \right]
\label{eq2c-21}  \, ,
\end{eqnarray}
\begin{eqnarray}
&&\hspace{-5mm}
A = 7\sigma^{4}+64 \sigma^{3} + 254\sigma^{2} +520\sigma +451
\label{eq2c-22}  \, , \\
&&\hspace{-5mm}
B = 3\sigma^{7} - 21 \sigma^{6} -582\sigma^{5} -4378\sigma^{4}-18589 \sigma^{3}   \nonumber  \\
&&\hspace{9mm}
- 48333\sigma^{2} -70688\sigma -44036
\label{eq2c-23} \, .
\end{eqnarray}
We also solved Eq.~(\ref{eq2c-19}) numerically and obtained the numerical solution $s_{num}(\sigma)$.  In Figs.~2(a) and 2(b), we plot $s(\sigma)$ and $P_{1}(s_{peak})$, respectively.  The dashed curve, dash-dotted curve, and solid curve correspond to $s_{1st}(\sigma)$, $s_{3rd}(\sigma)$ and $s_{num}(\sigma)$, respectively.  In Fig.~2(b), the solid curve and dash-dotted curve are indistinguishable.  It can be seen from Figs.~2(a) and 2(b) that the third-order approximation is sufficiently accurate for the present purposes.

%______________________________________________ Fig.\ 2
\begin{figure}
\begin{center}
\includegraphics[angle=0,width=0.48\textwidth]{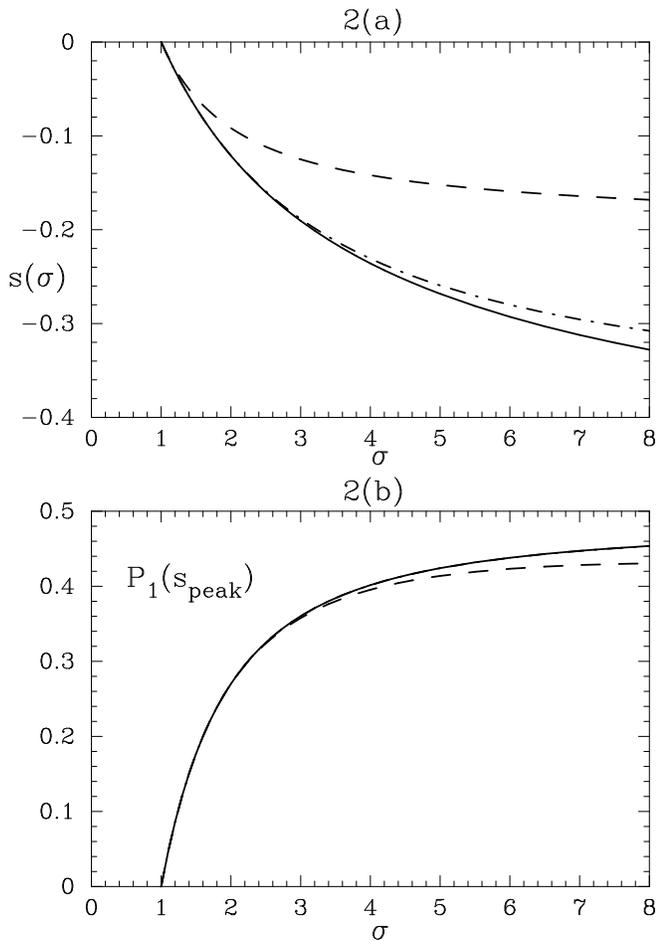}
\end{center}
\caption{Plotting of $s(\sigma)$ and $P_{1}(s_{peak})$.  Figures 2(a) and 2(b) are $s(\sigma)$ and $P_{1}(s_{peak})$, respectively.  The dashed curve, dash-dotted curve, and solid curve correspond to the first-order approximation, third-order approximation, and numerical solution, respectively.}
\end{figure}

\subsection{Scaling Law for Spectral Intensity Function}

  Let us now solve the rate equations of Eqs.~(\ref{eq2a-3}) and (\ref{eq2a-4}) with the result of Eq.~(\ref{eq2c-14}) for $P_{1}(s)$.  We consider the CMB photons for the initial distribution.  For the inverse Compton scattering by high-energy electrons, we are interested in high-energy spectrum such as X-rays ($\sim$ keV) and gamma-rays ($\sim$ MeV).  Therefore, one can safely assume
\begin{eqnarray}
&&\hspace{-10mm}
   x \equiv \frac{\omega}{k_{B}T_{CMB}} \gg 1
\label{eq2d-1}
\end{eqnarray}
for scattered photons.  For the $\gamma$-parameters, we assume the same condition used in the scaling law for $P_{1}(s)$, namely,
\begin{eqnarray}
&&\hspace{-10mm}
1 \ll \gamma_{min} \ll \gamma_{max}  \, .
\label{eq2d-2}
\end{eqnarray}
Under these assumptions, Eqs.~(\ref{eq2a-3}) and (\ref{eq2a-4}) are much simplified, and can be solved analytically.  The derivation is straightforward, however, it is lengthy.  Therefore, we give the derivation in Appendix B in detail, and quote the final results here.

  According to Eqs.~(\ref{eqB-38}) and (\ref{eqB-40}), one has the following results:
\begin{eqnarray}
&&\hspace{-5mm}
\frac{d I(X)}{d \tau}= I_0
\left[ 3(\sigma-1) X^3\int_{X}^{\infty} \frac{dt}{t} \frac{1}{e^t-1}
\right.
\nonumber \\
&&\hspace{-8mm}
\times
\left\{ -\frac{2}{\sigma+5}
+\frac{1}{\sigma+3}\left(\frac{\sigma-1}{\sigma+3}-2\ln\frac{t}{X}\right) \frac{t}{X} +\frac{1}{\sigma+1}\frac{t^{2}}{X^{2}} \right\}
\nonumber \\
&&\hspace{-8mm}
\left. + \, \frac{6(\sigma-1)(\sigma^2+4\sigma+11)}{(\sigma+1)(\sigma+3)^2(\sigma+5)}
\frac{1}{X^{(\sigma-1)/2}}
 \int_{0}^{X}dt\frac{t^{(\sigma+3)/2}}{e^t-1}
\right]
\label{eq2d-3}  \, ,  \\
&&\hspace{-5mm}
\frac{d n(X)}{d \tau} = \frac{1}{64 \gamma_{min}^{6}} \frac{1}{I_{0}X^{3}} \frac{d I(X)}{d \tau}  \, ,
\label{eq2d-4}  \\
&&\hspace{0mm}
X = \frac{x}{4 \gamma_{min}^{2}}
\label{eq2d-5}  \, ,
\end{eqnarray}
where $I_{0}=(k_{B}T_{CMB})^{3}/2\pi^{2}$.  It should be emphasized that the function $dI(X)/d\tau$ depends on $\gamma_{min}$ only through $X$.  Therefore, $dI(X)/d\tau$ has the scaling law.  On the other hand, the function $dn(X)/d\tau$ does not have the scaling law because of the factor 1/64$\gamma_{min}^{6}$ in Eq.~(\ref{eq2d-4}).

  For $X \gg 1$, Eq.~(\ref{eq2d-3}) is further simplified as follows:
\begin{eqnarray}
&&\hspace{-10mm}
\frac{d I(X)}{d \tau} = I_0
\frac{6(\sigma-1)(\sigma^2+4\sigma+11)}{(\sigma+1)(\sigma+3)^2(\sigma+5)}
\nonumber \\
&&\hspace{3mm}
\times \Gamma\left(\frac{\sigma+5}{2}\right)
\zeta\left(\frac{\sigma+5}{2}\right) X^{-(\sigma-1)/2}
\label{eq2d-6}  \, ,  \\
&&\hspace{-10mm}
\zeta(z) = \frac{1}{\Gamma(z)} \int_{0}^{\infty} dt \frac{t^{z-1}}{e^{t}-1}
\label{eq2d-7}  \, ,
\end{eqnarray}
where $\zeta(z)$ is the Riemann's zeta function.  We call Eq.~(\ref{eq2d-6}) the power-law approximation.

  In Fig.~3, we plot $dI(X)/d\tau$ of Eq.~(\ref{eq2d-3}) as a function of $X$ for typical $\sigma$-values for illustrative purposes.  The dashed curve, dash-dotted curve, and solid curve correspond to $\sigma$=2.5, 3.5, and 4.5, respectively.  The peak position and peak height depend only on the power-index parameter.  It should be noted that $dI(X)/d\tau \propto X$ for $X \ll 1$, and $dI(X)/d\tau \propto X^{-(\sigma-1)/2}$ for $X \gg 1$.  Therefore, the slope of the downward curves in Fig.~3 will determine the $\sigma$-value.

%______________________________________________ Fig.\ 3
\begin{figure}
\begin{center}
\includegraphics[angle=0,width=0.45\textwidth]{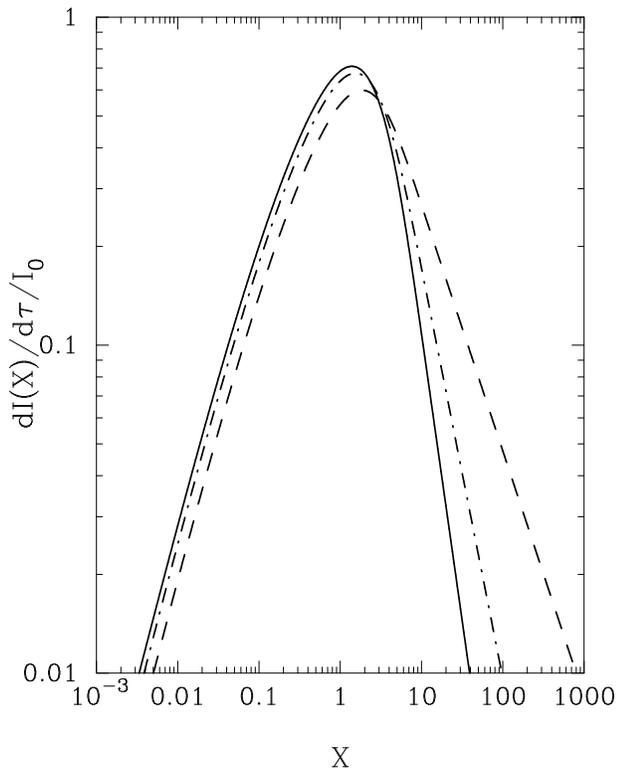}
\end{center}
\caption{Plotting of $dI(X)/d\tau$.  The dashed curve, dash-dotted curve, and solid curve correspond to $\sigma$ = 2.5, 3.5, and 4.5, respectively.}
\end{figure}

  In Figs.~4(a) and 4(b), we plot the peak position and peak height of the spectral intensity function as a function of $\sigma$, respectively.  The solid curves correspond to the numerical values.  The dash-dotted curves are the results of analytical fitting functions.  They are given by
\begin{eqnarray}
&&\hspace{-7mm}
X_{peak} = 1 + \frac{1}{\Sigma} \left(a_{0} + a_{1} \Sigma^{1/4} + a_{2} \Sigma^{1/2} \right) \, ,
\label{eq2d-8}  \\
&&\hspace{-10mm}
\frac{d I(X_{peak})}{d \tau I_{0}} = \frac{3}{4} \frac{ \Sigma \left(4 + 6 \Sigma + \Sigma^{2} \right) }{\left(b_{0} + b_{1} \Sigma + b_{2} \Sigma^{2} + \Sigma^{3} \right)}  \, ,
\label{eq2d-9}  \\
&&\hspace{1mm}
\Sigma \equiv \sigma - 1  \, .
\label{eq2d-10}
\end{eqnarray}
The fitting parameters are $a_{0}$=$-$2.18351, $a_{1}$=5.37131 and $a_{2}$=$-$2.02638 for the peak position, and $b_{0}$=2.60331, $b_{1}$=6.6352 and $b_{2}$=5.6526 for the peak height.  The errors of the fitting functions in the region $2 \le \sigma \le 10$ are less than 0.15\% and 0.10\% for $X_{peak}$ and $dI(X_{peak})/d\tau$, respectively.  In Fig.~4, two curves are totally indistinguishable.

%______________________________________________ Fig.\ 4
\begin{figure}
\begin{center}
\includegraphics[angle=0,width=0.5\textwidth]{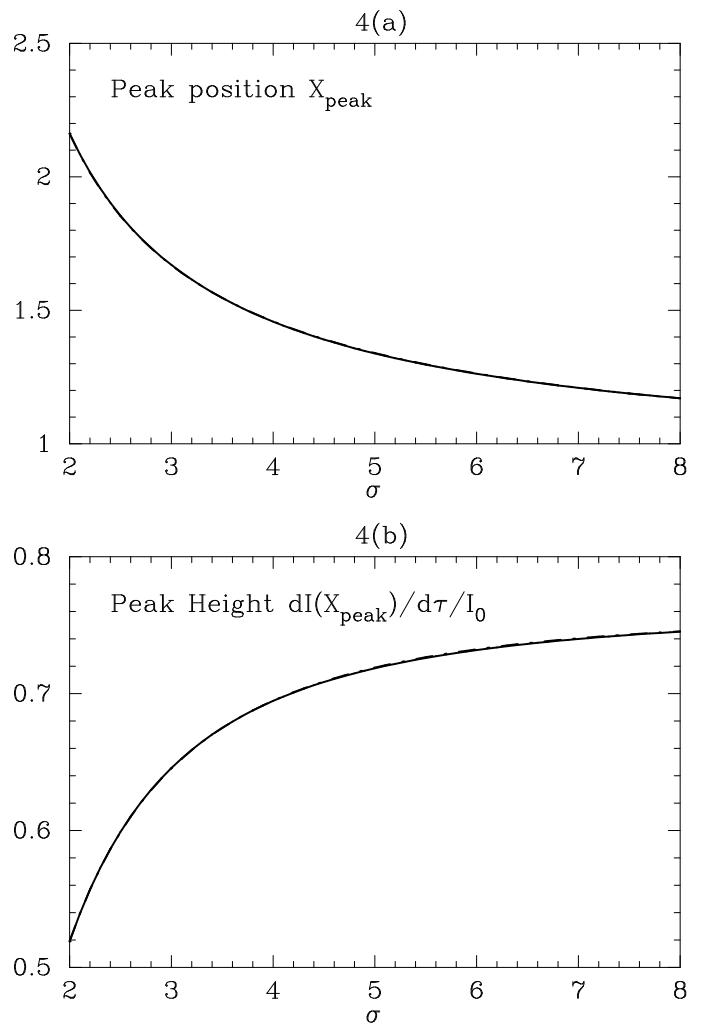}
\end{center}
\caption{Plotting of the peak position and peak height of the spectral intensity function as a function of $\sigma$.  The solid curves correspond to the numerical values.  The dash-dotted curves are the results of the analytical fitting functions.}
\end{figure}

  Before closing this section, let us compare the present result with the literature.  In the textbook by Rybicki and Lightman\cite{rybi79}, the expression for the power-law approximation is given by Eq.~(7.31).  The scaling law is hidden in the expression of Eq.~(7.31).  Inserting the explicit form of the normalization constant $C=n_{e}(\sigma-1) \gamma_{min}^{\sigma-1}$ in Eq.~(7.31), one finally obtains as follows:
\begin{eqnarray}
&&\hspace{-17mm}
\frac{d E}{dV d \tau d \epsilon_{1}} \propto I_{0} G(\sigma) X^{-(\sigma-1)/2}
\label{eq2d-11}  \, ,  \\
&&\hspace{-10mm}
G(\sigma) = \frac{6(\sigma-1)(\sigma^2+4\sigma+11)}{(\sigma+1)(\sigma+3)^2(\sigma+5)}  
\nonumber  \\
&&\hspace{0mm}
\times \Gamma\left(\frac{\sigma+5}{2}\right) \zeta\left(\frac{\sigma+5}{2}\right)
\label{eq2d-12}  \, ,
\end{eqnarray}
which agrees with Eq.~(\ref{eq2d-6}).  The applicable range of the power-law approximation of Eq.~(7.31) is given\cite{rybi79} by
\begin{eqnarray}
&&\hspace{-10mm}
4 \gamma_{min}^{2}  \ll  \frac{\epsilon_{1}}{\bar{\epsilon}}  \ll  4 \gamma_{max}^{2}
\label{eq2d-13}  \, ,
\end{eqnarray}
where $\bar{\epsilon}$ is a typical energy of initial photon distribution.  In the case of the CMB photon distribution, one can use $\bar{\epsilon}=k_{B}T_{CMB}$.  Therefore, one obtains the condition for $X$ as follows:
\begin{eqnarray}
&&\hspace{-10mm}
1  \ll  X  \ll  \frac{1}{R^{2}}
\label{eq2d-14}  \, ,
\end{eqnarray}
which again agrees with the condition of the present paper.  It is needless to mention that the full expression of Eq.~(\ref{eq2d-3}) has to be used for $X \leq O(1)$ as shown in Fig.~3.

\section{Astrophysical Applications of Scaling Laws}

\subsection{X-ray region}

  In the present section, we show an application of the scaling law.  Recently, observational studies on the inverse Compton scattering have been done quite extensively, for example, X-ray observations from radio galaxies with Chandra\cite{fabi03, erlu06}.

  In Fig.~5, we plot $dI(X)/d\tau$ as a function of the photon energy $\omega$ in X-ray energy region for a typical value $\sigma$=2.5.  Figures 5(a), 5(b), 5(c) and 5(d) correspond to $\gamma_{min}$=500, 1$\times10^{3}$, 2$\times10^{3}$ and 3$\times10^{3}$, respectively.  The solid curve corresponds to the full calculation of Eq.~(\ref{eq2d-3}).  The dash-dotted curve is the power-law approximation of Eq.~(\ref{eq2d-6}).  In Fig.~5, the peak height is independent of the $\gamma_{min}$-values as pointed in the last section.  On the other hand, the peak position is shifting toward to high-energy side as the $\gamma_{min}$-value increases.

%______________________________________________ Fig.\ 5
\begin{figure}
\begin{center}
\includegraphics[angle=0,width=0.5\textwidth]{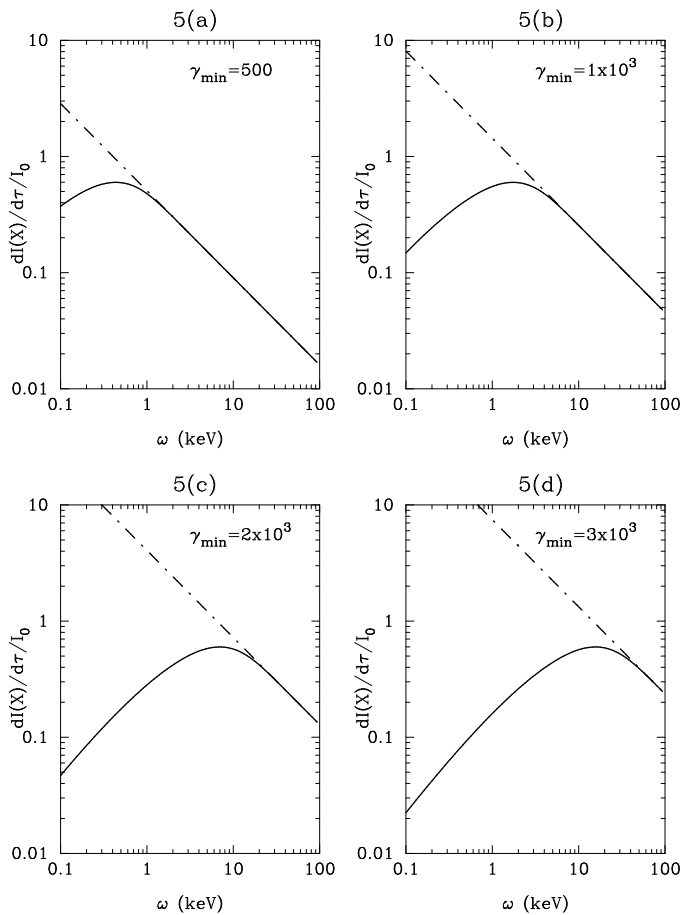}
\end{center}
\caption{Plotting of $dI(X)/d\tau$ as a function of the photon energy $\omega$ in X-ray energy regions for a typical value $\sigma$=2.5.  Figures 5(a), 5(b), 5(c) and 5(d) correspond to $\gamma_{min}$=500, 1$\times10^{3}$, 2$\times10^{3}$ and 3$\times10^{3}$, respectively.  The solid curve corresponds to the full calculation of Eq.~(\ref{eq2d-3}).  The dash-dotted curve is the power-law approximation of Eq.~(\ref{eq2d-6}).}
\end{figure}

  By measuring the slope of the downward curve in Fig.~5, one can determine the $\sigma$-value, because
\begin{eqnarray}
&&\hspace{-10mm}
\frac{d I(\omega)}{d \tau} \propto \omega^{-(\sigma-1)/2}
\label{eq3a-1}
\end{eqnarray}
is valid.  One can also determine the $\sigma$-value by measuring the peak height in Fig.~5 with the expression of Eq.~(\ref{eq2d-9}).  This will serve as an independent check for the $\sigma$-value.  On the other hand, the $\gamma_{min}$-value is determined by measuring the peak position $\omega_{peak}$ in Fig.~5.  Using the relation of $\omega_{peak}$ with $X_{peak}$, namely,
\begin{eqnarray}
&&\hspace{-10mm}
X_{peak} = \frac{1}{4 \gamma_{min}^{2}} \frac{\omega_{peak}}{k_{B}T_{CMB}}   \, ,
\label{eq3a-2}
\end{eqnarray}
the $\gamma_{min}$-value is determined by
\begin{eqnarray}
&&\hspace{-10mm}
\gamma_{min} = \left[\frac{1}{4 X_{peak}} \frac{\omega_{peak}}{k_{B}T_{CMB}} \right]^{1/2}  \, ,
\label{eq3a-3}
\end{eqnarray}
where $X_{peak}$ is calculated by the RHS of Eq.~(\ref{eq2d-8}) with the measured $\sigma$-value.  It can be seen from Fig.~5 that the X-ray observations have sensitivities to $\gamma_{min}$=500 $\sim$ 3$\times 10^{3}$ region. 

  Before closing this subsection, let us study the applicability of the power-law approximation used in the literature.  The condition for the power-law approximation $X \gg 1$ reads
\begin{eqnarray}
&&\hspace{-10mm}
\omega \gg 4 \gamma_{min}^{2} k_{B}T_{CMB}
\label{eq3a-4}  \, .
\end{eqnarray}
In the case of $\gamma_{min}=1\times10^{3}$, for example, one has $\omega \gg 1$ keV.  As shown in Fig.~5(b), the error of the power-law approximation is quite large in $\omega \sim O(1)$ keV region.

\subsection{gamma-ray region}

  With the scaling law for the spectral intensity function, one can extend the present formalism to the gamma-ray region.  In Fig.~6, we plot the same figure as Fig.~5 for the gamma-ray region.  Because of the scaling law, the factor $\sqrt{1000}(\approx 31.6)$ should be simply multiplied to the $\gamma_{min}$-values of the keV region in order to obtain the spectral intensity function in the MeV region.  Therefore, observations in this energy region will have sensitivities to $\gamma_{min}$=16$\times 10^{3}$ $\sim$ 95$\times 10^{3}$ region.  Similarly, the factor 1000 should be multiplied to the $\gamma_{min}$-values of the keV region in order to obtain the parameter values in the GeV region.

%______________________________________________ Fig.\ 6
\begin{figure}
\begin{center}
\includegraphics[angle=0,width=0.5\textwidth]{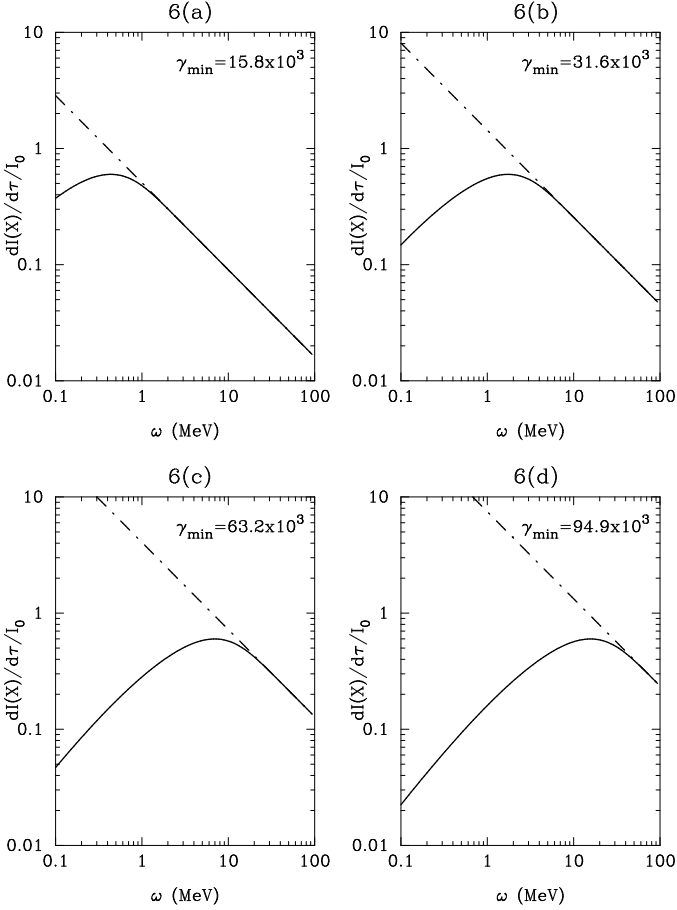}
\end{center}
\caption{Plotting of $dI(X)/d\tau$ as a function of the photon energy $\omega$ in gamma-ray energy regions for a typical value $\sigma$=2.5.  Figures 6(a), 6(b), 6(c) and 6(d) correspond to $\gamma_{min}$=15.8$\times10^{3}$, 31.6$\times10^{3}$, 63.2$\times10^{3}$ and 94.9$\times10^{3}$, respectively.  The solid curve corresponds to the full calculation of Eq.~(\ref{eq2d-3}).  The dash-dotted curve is the power-law approximation of Eq.~(\ref{eq2d-6}).}
\end{figure}

\section{Concluding Remarks}

  In the NK paper\cite{noza09a}, we derived the frequency redistribution function $P(s,\beta)$ for a frequency shift $s$ and electron velocity $\beta$.  The form was derived in the Thomson approximation, however, it was mathematically equivalent to that in the covariant formalism\cite{itoh98}.  Therefore the frequency redistribution function can be applicable from nonrelativistic electrons to extreme-relativistic electrons.
  
  In the present paper, we have extended the formalism to extreme-relativistic electrons.  First, we derived the analytic expression for $P(s,\gamma)$ in the approximation $\gamma \gg 1$.  It has been found that the present formalism is equivalent to Jones's formalism\cite{jone68}.

  By averaging $P(s,\gamma)$ over the nonthermal electron distribution function, we have calculated the probability distribution function $P_{1}(s)$.  As for the nonthermal distribution function, we have adopted a standard power-law distribution function of three parameters: the power index $\sigma$, minimum value $\gamma_{min}$, and maximum value $\gamma_{max}$ of the distribution range.  For the case $\gamma_{min} \gg 1$, we have found a scaling law in $P_{1}(s)$, where the peak position depends on $s- 2\ln 2\gamma_{min}$, and the peak height depends only on the power index parameter $\sigma$.

  We have calculated the spectral intensity function.  For the case of high-energy photons of $x \gg 1$, we have found a scaling law in $dI(x)/d\tau$, where the function depends on a new variable $X=x/(4\gamma_{min}^{2})$.  The peak position and peak height depend only on the power index parameter $\sigma$.  The $\gamma_{min}$-dependence of $dI(X)/d\tau$ is included in the variable $X$.

  We have applied the present formalism to the observation of the spectral intensity function in the X-ray and gamma-ray energy regions.  It has been found that the sensitivities of the observation in the X-ray and gamma-ray regions are $\gamma_{min}$=500 $\sim$ 3$\times 10^{3}$ and $\gamma_{min}$=16$\times 10^{3}$ $\sim$ 95$\times 10^{3}$, respectively.

  Finally, we have studied the applicability of the power-law approximation used in the literature.  In the case of $\gamma_{min}=1\times10^{3}$, for example, the error of the power-law approximation is quite large in $\omega \sim O(1)$ keV region.

\begin{acknowledgments}
This work is financially supported in part by the Grant-in-Aid of Japanese Ministry of Education, Culture, Sports, Science, and Technology under the contract \#21540277.  We would like to thank our referee for valuable suggestions.
\end{acknowledgments}

\appendix

\section{Comparison with Jones's Formalism}

  The double differential cross sections for extreme-relativistic electrons are given by Eqs.~(38) and (40) of Jones's paper\cite{jone68} as follows:
\begin{eqnarray}
&&\hspace{-10mm}
\frac{d^2N}{dtd\alpha} = \frac{2\pi r_{0}^{2}c}{\alpha_{1} \gamma^2}
\Biggl[2q \ln q + (1+2q)(1-q) \Biggr]
\label{eqA-1} \, ,  \\
&&\hspace{-10mm}
\frac{d^2N}{dtd\alpha} = \frac{\pi r_{0}^{2}c}{2\alpha_{1} \gamma^4}
\left[(q^{\prime}-1)\left(1+\frac{2}{q^{\prime}}\right) - 2 \ln q^{\prime} \right]
\label{eqA-2} \, .
\end{eqnarray}
The variables in Eqs.~(\ref{eqA-1}) and (\ref{eqA-2}) are related to the variables of the present paper as follows: $\sigma_T = 8\pi r_0^2/3$, $\alpha/\alpha_1 = e^{-s}$, $q = e^{-s}/4\gamma^2$, $q^{\prime}=4\gamma^2e^{-s}$, $\alpha = \theta_{CMB}x$, where $\theta_{CMB} = k_{B}T_{CMB}/mc^{2}$.  With these variables, Eqs.~(\ref{eqA-1}) and (\ref{eqA-2}) are rewritten as follows:
\begin{eqnarray}
&&\hspace{-15mm}
\frac{d^2N}{dtdx} = \frac{3\sigma_Tc}{32\gamma^4}\frac{1}{x}e^{-3s}
\Bigl[-\frac{1}{\gamma^2}+2e^s+8\gamma^2e^{2s}
\nonumber  \\
&&\hspace{+15mm}
-4\left(\lambda_{\gamma}+s\right)e^s \Bigr]
\label{eqA-3} \, ,  \\
&&\hspace{-15mm}
\frac{d^2N}{dtdx} = \frac{3\sigma_Tc}{32\gamma^4}\frac{1}{x}e^{-3s}
\Bigl[ 8\gamma^2e^s
+2e^{2s}
\nonumber  \\
&&\hspace{+15mm}
-4\left(\lambda_{\gamma}-s\right)e^{2s}
-\frac{1}{\gamma^2}e^{3s} \Bigr]
\label{eqA-4} \, .
\end{eqnarray}

  Let us denote the photon distribution function in Jones's formalism as $n_{J}(\alpha_1)$.  Then one has
\begin{eqnarray}
&&\hspace{-10mm}
n_{J}(\alpha_1)d\alpha_1 = \frac{m_e^3}{\pi^2(\hbar c)^3}
\frac{\alpha_1^2}{e^{\alpha_1/\theta_{CMB}}-1}d\alpha_1
\nonumber \\
&&\hspace{+8mm}
= \frac{(k_BT_{CMB})^3}{\pi^2(\hbar c)^3}x^3e^{3s} n(e^{s}x)ds
\label{eqA-5}  \, ,
\end{eqnarray}
where $n(e^{s}x) = 1/(e^{e^{s}x} - 1)$.  Averaging Eqs.~(\ref{eqA-3}) and (\ref{eqA-4}) over the photon momentum with the distribution function, one finally obtains
\begin{eqnarray}
&&\hspace{-15mm}
\int \frac{d^2N}{dtdx}n_J(\alpha_1)d\alpha_1 = \frac{(k_BT_{CMB})^3}{\pi^2(\hbar c)^3}\sigma_T c x^2
\nonumber  \\
&&\hspace{+15mm}
\times \int ds P_J(s,\gamma) n(e^{s}x)
\label{eqA-6}  \, ,
\end{eqnarray}
where the redistribution function in Jones's formalism is given by
\begin{eqnarray}
&&\hspace{-15mm}
P_{J}(s,\gamma) = \frac{3}{32\gamma^4}
\Bigl[-\frac{1}{\gamma^2}+2e^s+8\gamma^2e^{2s}
\nonumber  \\
&&\hspace{+15mm}
-4\left(\lambda_{\gamma}+s\right)e^s \Bigr]
\label{eqA-7} \, ,  \\
&&\hspace{-15mm}
P_{J}(s,\gamma) = \frac{3}{32\gamma^4}\Bigl[ 8\gamma^2e^s
+2e^{2s}
\nonumber  \\
&&\hspace{+15mm}
-4\left(\lambda_{\gamma}-s\right)e^{2s}
-\frac{1}{\gamma^2}e^{3s} \Bigr]
\label{eqA-8} \, .
\end{eqnarray}
Comparing Eqs.~(\ref{eqA-7}) and (\ref{eqA-8}) with Eqs.~(\ref{eq2b-13}) and (\ref{eq2b-14}), respectively, one finds
\begin{eqnarray}
&&\hspace{-10mm}
P(s,\gamma) = P_{J}(s,\gamma)
\label{eqA-9}  \, ,
\end{eqnarray}
which shows the equivalence of the two formalisms for extreme-relativistic electrons.

\section{Derivation of Equations (\ref{eq2d-3}) and (\ref{eq2d-4}) }

  We assume $x \gg 1$ for the scattered photons, and $1 \ll \gamma_{min} \ll \gamma_{max}$ for the $\gamma$-parameters.  Let us first solve Eq.~(\ref{eq2a-3}).  It can be rewritten as follows:
\begin{eqnarray}
&&\hspace{-5mm}
\frac{\partial n(x)}{\partial\tau} =
\nonumber  \\
&&\hspace{5mm}
\int_{-2\ln2\gamma_{min}}^{2\ln\gamma_{max}/\gamma_{min}}
dsP_{C}\left(s,\frac{\gamma_{min}}{\gamma_{max}}\right)
n\left(e^s4\gamma_{min}^2x\right)
\nonumber \\
&&\hspace{0mm}
+ \int_{-2\ln\gamma_{max}/\gamma_{min}}^{2\ln2\gamma_{min}}
dsP_{IC}\left(s,\frac{\gamma_{min}}{\gamma_{max}}\right)
n\left(e^s\frac{x}{4\gamma_{min}^2}\right)
\nonumber  \\
&&\hspace{0mm}
-n(x) \, .
\label{eqB-1}
\end{eqnarray}
Then we introduce the following new functions:
\begin{eqnarray}
&&\hspace{-10mm}
 \int_{-2\ln2\gamma_{min}}^{2\ln 1/R}
dsP_{C}\left(s,R\right)
n\left(e^{s}4\gamma_{min}^2x\right)
\nonumber  \\
&&\hspace{-5mm}
 \equiv n_1(x) + n_2(x)  \, ,
\label{eqB-2}
\end{eqnarray}
\begin{eqnarray}
&&\hspace{-10mm}
 \int_{-2\ln 1/R}^{2\ln2\gamma_{min}}
dsP_{IC}\left(s,R\right)
n\left(e^{s}\frac{x}{4\gamma_{min}^2}\right)
\nonumber  \\
&&\hspace{-5mm}
\equiv n_3(x) + n_4(x) \, ,
\label{eqB-3}
\end{eqnarray}
where $R \equiv \gamma_{min}/\gamma_{max}$.  In Eqs.~(\ref{eqB-2}) and (\ref{eqB-3}), the functions $n_{1}(x)$, ..., $n_{4}(x)$ are expressed as follows:
\begin{eqnarray}
&&\hspace{-10mm}
n_1(x) =
 \int_{-2\ln2\gamma_{min}}^{0}
dsP_{C}\left(s,0\right)
n\left(e^{s}Y\right)  \, ,
\label{eqB-4} \\
&&\hspace{-10mm}
n_2(x) =
 \int_{0}^{\infty}
dsP_{C}\left(s,0\right)
n\left(e^{s}Y\right)  \, ,
\label{eqB-5} \\
&&\hspace{-10mm}
n_3(x) = \int_{-\infty}^{0}
dsP_{IC}\left(s,0\right)
n\left(e^{s}X\right)  \, ,
\label{eqB-6} \\
&&\hspace{-10mm}
n_4(x) = \int_{0}^{2\ln2\gamma_{min}}
dsP_{IC}\left(s,0\right)
n\left(e^{s}X\right)  \, .
\label{eqB-7}
\end{eqnarray}
In deriving Eqs.~(\ref{eqB-4})--(\ref{eqB-7}), we put $R=0$ and used new variables:
\begin{eqnarray}
&&\hspace{-10mm}
X \equiv \frac{x}{4\gamma_{min}^2}  \, ,
\label{eqB-8}  \\
&&\hspace{-10mm}
Y \equiv 4\gamma_{min}^{2} x  \, .
\label{eqB-9}
\end{eqnarray}
Introducing $t = e^{s}Y$ into Eqs.~(\ref{eqB-4}) and (\ref{eqB-5}) and $t = e^{s}X$ into Eqs.~(\ref{eqB-6}) and (\ref{eqB-7}), and inserting the explicit forms of Eqs.~(\ref{eq2c-9})--(\ref{eq2c-12}), one obtains as follows:
\begin{eqnarray}
&&\hspace{-10mm}
n_{1}(x) = 3(\sigma-1) \frac{1}{Y} \int_{x}^{Y} dt \Bigl\{
-\frac{2}{\sigma+5}\frac{t^{2}}{Y^{2}}
\nonumber  \\
&&\hspace{-5mm}
+ \frac{1}{\sigma+3}\left(\frac{\sigma-1}{\sigma+3}+2\ln\frac{t}{Y}\right)\frac{t}{Y} + \frac{1}{\sigma+1} \Bigr\} n(t)  \, ,
\label{eqB-10}  \\
&&\hspace{-10mm}
n_{2}(x) = \frac{6(\sigma-1)(\sigma^2+4\sigma+11)}{(\sigma+1)(\sigma+3)^2(\sigma+5)} Y^{(\sigma-1)/2}
\nonumber  \\
&&\hspace{0mm}
\times \int_{Y}^{\infty}dt t^{-(\sigma+1)/2} n(t)  \, ,
\label{eqB-11}
\end{eqnarray}
\begin{eqnarray}
&&\hspace{-10mm}
n_{3}(x) = \frac{6(\sigma-1)(\sigma^2+4\sigma+11)}{(\sigma+1)(\sigma+3)^2(\sigma+5)} \frac{1}{x^{3}} \frac{1}{X^{(\sigma-1)/2}}
\nonumber  \\
&&\hspace{0mm}
\times \int_{0}^{X}dt \, t^{(\sigma+3)/2} n(t)  \, ,
\label{eqB-12}  \\
&&\hspace{-10mm}
n_{4}(x) = 3(\sigma-1) \frac{X^{3}}{x^{3}} \int_{X}^{x} \frac{dt}{t} \Bigl\{
-\frac{2}{\sigma+5}
\nonumber  \\
&&\hspace{-10mm}
+ \frac{1}{\sigma+3}\left(\frac{\sigma-1}{\sigma+3}-2\ln\frac{t}{X}\right)\frac{t}{X} + \frac{1}{\sigma+1} \frac{t^{2}}{X^{2}} \Bigr\} n(t)  \, .
\label{eqB-13}
\end{eqnarray}

  Now let us consider the CMB photon distribution function
\begin{eqnarray}
&&\hspace{-20mm}
n_{0}(t) = \frac{1}{e^t-1}
\label{eqB-14}
\end{eqnarray}
for the initial distribution.  Inserting Eq.~(\ref{eqB-14}) into Eqs.~(\ref{eqB-10})--(\ref{eqB-13}), one has for $x \gg 1$
\begin{eqnarray}
&&\hspace{-10mm}
n_{1}(x) = 0  \, ,
\label{eqB-15}  \\
&&\hspace{-10mm}
n_{2}(x) = 0  \, ,
\label{eqB-16}
\end{eqnarray}
\begin{eqnarray}
&&\hspace{-10mm}
n_{3}(x) = \frac{6(\sigma-1)(\sigma^2+4\sigma+11)}{(\sigma+1)(\sigma+3)^2(\sigma+5)}
\frac{1}{x^{3}} \frac{1}{X^{(\sigma-1)/2}}
\nonumber  \\
&&\hspace{0mm}
\times \int_{0}^{X}dt \, t^{(\sigma+3)/2} \frac{1}{e^{t}-1}  \, ,
\label{eqB-17}  \\
&&\hspace{-10mm}
n_{4}(x) = 3(\sigma-1) \frac{X^{3}}{x^{3}} \int_{X}^{\infty} \frac{dt}{t} \frac{1}{e^{t}-1} \Bigl\{
-\frac{2}{\sigma+5}
\nonumber  \\
&&\hspace{-5mm}
+ \frac{1}{\sigma+3}\left(\frac{\sigma-1}{\sigma+3}-2\ln\frac{t}{X}\right)\frac{t}{X} + \frac{1}{\sigma+1}\frac{t^{2}}{X^{2}} \Bigr\}  \, .
\label{eqB-18}
\end{eqnarray}
The last term in Eq.~(\ref{eqB-1}) can be safely neglected for $x \gg 1$.  Therefore, one obtains
\begin{eqnarray}
&&\hspace{-12mm}
\frac{\partial n(x)}{\partial\tau}
= n_{3}(x) + n_{4}(x)  \, .
\label{eqB-19}
\end{eqnarray}
One finally obtains
\begin{eqnarray}
&&\hspace{-10mm}
\frac{d n(X)}{d \tau} = \frac{1}{x^{3}}
\left[ 3(\sigma-1) X^3\int_{X}^{\infty} \frac{dt}{t} \frac{1}{e^t-1}
\right.
\nonumber \\
&&\hspace{-12mm}
\times
\left\{ -\frac{2}{\sigma+5}
+\frac{1}{\sigma+3}\left(\frac{\sigma-1}{\sigma+3}-2\ln\frac{t}{X}\right) \frac{t}{X} +\frac{1}{\sigma+1}\frac{t^{2}}{X^{2}} \right\}
\nonumber \\
&&\hspace{-12mm}
\left. + \, \frac{6(\sigma-1)(\sigma^2+4\sigma+11)}{(\sigma+1)(\sigma+3)^2(\sigma+5)}
\frac{1}{X^{(\sigma-1)/2}}
 \int_{0}^{X}dt\frac{t^{(\sigma+3)/2}}{e^t-1}
\right]
\label{eqB-20}  \, .
\end{eqnarray}

  One can also solve Eq.~(\ref{eq2a-4}) in a similar manner.  It can be rewritten as follows:
\begin{eqnarray}
&&\hspace{-5mm}
\frac{\partial I(x)}{\partial\tau}
= 
\nonumber \\
&&\hspace{5mm}
\int_{-2\ln2\gamma_{min}}^{2\ln\gamma_{max}/\gamma_{min}}
dsP_{C}\left(s,\frac{\gamma_{min}}{\gamma_{max}}\right)
I\left(e^{-s}\frac{x}{4\gamma_{min}^2}\right)
\nonumber \\
&&\hspace{0mm}
+ \int_{-2\ln\gamma_{max}/\gamma_{min}}^{2\ln2\gamma_{min}}
dsP_{IC}\left(s,\frac{\gamma_{min}}{\gamma_{max}}\right)
I\left(e^{-s}{4\gamma_{min}^2}x\right)
\nonumber  \\
&&\hspace{0mm}
-I(x) \, .
\label{eqB-21}
\end{eqnarray}
Then we introduce the following new functions:
\begin{eqnarray}
&&\hspace{-10mm}
 \int_{-2\ln2\gamma_{min}}^{2\ln 1/R}
dsP_{C}\left(s,R\right)
I\left(e^{-s}\frac{x}{4\gamma_{min}^2}\right)
\nonumber  \\
&&\hspace{-5mm}
 \equiv I_1(x) + I_2(x)  \, ,
\label{eqB-22}
\end{eqnarray}
\begin{eqnarray}
&&\hspace{-10mm}
 \int_{-2\ln 1/R}^{2\ln2\gamma_{min}}
dsP_{IC}\left(s,R\right)
I\left(e^{-s}{4\gamma_{min}^2}x\right)
\nonumber  \\
&&\hspace{-5mm}
\equiv I_3(x) + I_4(x) \, .
\label{eqB-23}
\end{eqnarray}
In Eqs.~(\ref{eqB-22}) and (\ref{eqB-23}), the functions $I_{1}(x)$, ..., $I_{4}(x)$ are expressed as follows:
\begin{eqnarray}
&&\hspace{-10mm}
I_1(x) =
 \int_{-2\ln2\gamma_{min}}^{0}
dsP_{C}\left(s,0\right)
I\left(e^{-s}X\right)  \, ,
\label{eqB-24} \\
&&\hspace{-10mm}
I_2(x) =
 \int_{0}^{\infty}
dsP_{C}\left(s,0\right)
I\left(e^{-s}X\right)  \, ,
\label{eqB-25} \\
&&\hspace{-10mm}
I_3(x) = \int_{-\infty}^{0}
dsP_{IC}\left(s,0\right)
I\left(e^{-s}Y\right)  \, ,
\label{eqB-26} \\
&&\hspace{-10mm}
I_4(x) = \int_{0}^{2\ln2\gamma_{min}}
dsP_{IC}\left(s,0\right)
I\left(e^{-s}Y\right)  \, ,
\label{eqB-27}
\end{eqnarray}
where we put $R=0$ and used the variables $X$ and $Y$.  Introducing $t = e^{-s}X$ into Eqs.~(\ref{eqB-24}) and (\ref{eqB-25}) and $t = e^{-s}Y$ into Eqs.~(\ref{eqB-26}) and (\ref{eqB-27}), and inserting the explicit forms of Eqs.~(\ref{eq2c-9})--(\ref{eq2c-12}), one obtains as follows:
\begin{eqnarray}
&&\hspace{-10mm}
I_{1}(x) = 3(\sigma-1) X^{3} \int_{X}^{x} \frac{dt}{t^{4}} \Bigl\{
-\frac{2}{\sigma+5}
\nonumber  \\
&&\hspace{-10mm}
+ \frac{1}{\sigma+3}\left(\frac{\sigma-1}{\sigma+3}-2\ln\frac{t}{X}\right)\frac{t}{X} + \frac{1}{\sigma+1} \frac{t^{2}}{X^{2}} \Bigr\} I(t)  \, ,
\label{eqB-28}  \\
&&\hspace{-10mm}
I_{2}(x) = \frac{6(\sigma-1)(\sigma^2+4\sigma+11)}{(\sigma+1)(\sigma+3)^2(\sigma+5)}
\frac{1}{X^{(\sigma-1)/2}} 
\nonumber  \\
&&\hspace{0mm}
\times \int_{0}^{X}dt \, t^{(\sigma-3)/2} I(t)  \, ,
\label{eqB-29}
\end{eqnarray}
\begin{eqnarray}
&&\hspace{-10mm}
I_{3}(x) = \frac{6(\sigma-1)(\sigma^2+4\sigma+11)}{(\sigma+1)(\sigma+3)^2(\sigma+5)} x^{3} Y^{(\sigma-1)/2}
\nonumber  \\
&&\hspace{0mm}
\times \int_{Y}^{\infty}dt \, t^{-(\sigma+7)/2} I(t)  \, ,
\label{eqB-30}  \\
&&\hspace{-10mm}
I_{4}(x) = 3(\sigma-1) \frac{x^{3}}{Y} \int_{x}^{Y} \frac{dt}{t^{3}} \Bigl\{
-\frac{2}{\sigma+5}\frac{t^{2}}{Y^{2}}
\nonumber  \\
&&\hspace{-7mm}
+ \frac{1}{\sigma+3}\left(\frac{\sigma-1}{\sigma+3}+2\ln\frac{t}{Y}\right)\frac{t}{Y} + \frac{1}{\sigma+1} \Bigr\} I(t)  \, .
\label{eqB-31}
\end{eqnarray}

  Now let us consider the CMB photon distribution function
\begin{eqnarray}
&&\hspace{-20mm}
I_{0}(t) = I_0 t^{3} n_{0}(t)
\label{eqB-32}
\end{eqnarray}
for the initial distribution, where $I_0 = (k_BT_{CMB})^3/2\pi^2$ and $n_{0}(t)$ is given by Eq.~(\ref{eqB-14}).  Inserting Eq.~(\ref{eqB-32}) into Eqs.~(\ref{eqB-28})--(\ref{eqB-31}), one has for $x \gg 1$
\begin{eqnarray}
&&\hspace{-10mm}
I_{1}(x) = 3I_{0}(\sigma-1) X^{3} \int_{X}^{\infty} \frac{dt}{t} \frac{1}{e^{t}-1} \Bigl\{
-\frac{2}{\sigma+5}
\nonumber  \\
&&\hspace{-5mm}
+ \frac{1}{\sigma+3}\left(\frac{\sigma-1}{\sigma+3}-2\ln\frac{t}{X}\right)\frac{t}{X} + \frac{1}{\sigma+1}\frac{t^{2}}{X^{2}} \Bigr\}  \, ,
\label{eqB-33}  \\
&&\hspace{-10mm}
I_{2}(x) = I_{0}\frac{6(\sigma-1)(\sigma^2+4\sigma+11)}{(\sigma+1)(\sigma+3)^2(\sigma+5)}
\frac{1}{X^{(\sigma-1)/2}} 
\nonumber  \\
&&\hspace{0mm}
\times \int_{0}^{X}dt \, t^{(\sigma+3)/2} \frac{1}{e^{t}-1}  \, ,
\label{eqB-34}
\end{eqnarray}
\begin{eqnarray}
&&\hspace{-10mm}
I_{3}(x) = 0  \, ,
\label{eqB-35}  \\
&&\hspace{-10mm}
I_{4}(x) = 0  \, .
\label{eqB-36}
\end{eqnarray}
The last term in Eq.~(\ref{eqB-21}) can be safely neglected for $x \gg 1$.  Therefore, one obtains
\begin{eqnarray}
&&\hspace{-12mm}
\frac{\partial I(x)}{\partial\tau}
= I_{1}(x) + I_{2}(x)  \, .
\label{eqB-37}
\end{eqnarray}
One finally obtains
\begin{eqnarray}
&&\hspace{-10mm}
\frac{d I(X)}{d \tau} = I_{0}
\left[ 3(\sigma-1) X^3\int_{X}^{\infty} \frac{dt}{t} \frac{1}{e^t-1}
\right.
\nonumber \\
&&\hspace{-12mm}
\times
\left\{ -\frac{2}{\sigma+5}
+\frac{1}{\sigma+3}\left(\frac{\sigma-1}{\sigma+3}-2\ln\frac{t}{X}\right) \frac{t}{X} +\frac{1}{\sigma+1}\frac{t^{2}}{X^{2}} \right\}
\nonumber \\
&&\hspace{-12mm}
\left. + \, \frac{6(\sigma-1)(\sigma^2+4\sigma+11)}{(\sigma+1)(\sigma+3)^2(\sigma+5)}
\frac{1}{X^{(\sigma-1)/2}}
 \int_{0}^{X}dt\frac{t^{(\sigma+3)/2}}{e^t-1}
\right]
\label{eqB-38}  \, .
\end{eqnarray}
Comparing Eq.~(\ref{eqB-38}) with Eq.~(\ref{eqB-20}), one has the following relation:
\begin{eqnarray}
&&\hspace{-10mm}
\frac{d n(X)}{d \tau} = \frac{1}{I_{0}x^{3}} \frac{d I(X)}{d \tau}  \, ,
\label{eqB-39}  \\
&&\hspace{-1mm}
= \frac{1}{64 \gamma_{min}^{6}} \frac{1}{I_{0} X^{3}} \frac{d I(X)}{d \tau}  \, .
\label{eqB-40}
\end{eqnarray}

%\newpage %Just because of unusual number of tables stacked at end

\bibliography{apssamp}% Produces the bibliography via BibTeX.

\end{document}